\documentclass[twocolumn]{revtex4}

\usepackage{graphicx}
\begin{document}

\title{Electric Field Effect Thermoelectric Transport in Individual Silicon and Germanium/Silicon Nanowires}

\author{Yuri M. Brovman$^1$}
\author{Joshua P. Small$^1$}
\author{Yongjie Hu$^2$}
\author{Ying Fang$^2$}
\author{Charles M. Lieber$^2$}
\author{Philip Kim$^1$}
\affiliation{$^1$Department of Applied Physics and Applied Mathematics and Department of Physics, Columbia University, New York, New York, 10027, USA}
\affiliation{$^2$Department of Chemistry and Chemical Biology, Harvard University, Cambridge, MA 02139, USA}

\begin{abstract}
We have simultaneously measured conductance and thermoelectric power (TEP) of individual silicon and germanium/silicon core/shell nanowires in the field effect transistor device configuration. As the applied gate voltage changes, the TEP shows distinctly different behaviors while the electrical conductance exhibits the turn-off, subthreshold, and saturation regimes respectively. At room temperature, peak TEP value of $\sim 300~\mu$V/K is observed in the subthreshold regime of the Si devices. The temperature dependence of the saturated TEP values are used to estimate the carrier doping of Si nanowires.
\end{abstract}

\maketitle

The electronic properties of Si and Ge/Si nanowires (NW) have attracted considerable attention for applications in next generation field effect transistors (FET)~\cite{Si_schottky,Si_growth,Si_dopant,Si_FET1,Si_mass,Si_FET2,GeSi_PNAS,GeSi_FET,GeSi_theory,GeSi_2THz}. Unlike carbon nanotubes, Si based NWs can be synthesized with controlled diameters and doping levels for rational device design. Thermal and thermoelectric transport properties of these NWs are also of interest for potential use in thermoelectric power conversion applications. Thermal transport studies in Si NWs have shown that increasing the surface roughness ~\cite{SiNW_ZT1} or the enhanced phonon-drag~\cite{SiNW_ZT1,SiNW_ZT2} can increase thermoelectric efficiency in Si NWs. The doping level of individual NWs can often be adjustable using the electric field effect (EFE) using the gate electrode in a field effect transistor (FET) device configuration. Since electronic properties of the NWs are sensitively dependent on the carrier density, TEP thus can be adjusted by the EFE. For NWs made of narrow gap semiconductors, such PbSe, Sb$_2$Te$_3$, the EFE modulated TEP has been measured in the FET device configuration where a large modulation of TEP has been demonstrated electrical control of thermoelectric efficiency~\cite{PbSe_FET, SbTe_TEP, InAs_TEP}. Recently, gate-modulated TEP, $S$, and electric conductance, $\sigma$, has also been measured in Ge/Si core-shell heterostructured NWs, where the optimization of the power factor, $\sigma S^2$ has been demonstrated employing the EFE~\cite{GeSi_TEP}. Extending the EFE modulated conductivity and TEP measurements on individual Si NWs may provide a new insight to understand the carrier density dependence of thermoelectric transport properties in Si based semiconducting nanostructures, which can be employed for optimizing their applicability in thermoelectric applications. However, the relatively large channel impedance in the Si NW devices, especially near the turn-off regime of the devices poses the experimental challenge of measuring thermoelectric signal simultaneously with the electrical conduction.

\begin{figure}
\includegraphics[width=1.0\linewidth]{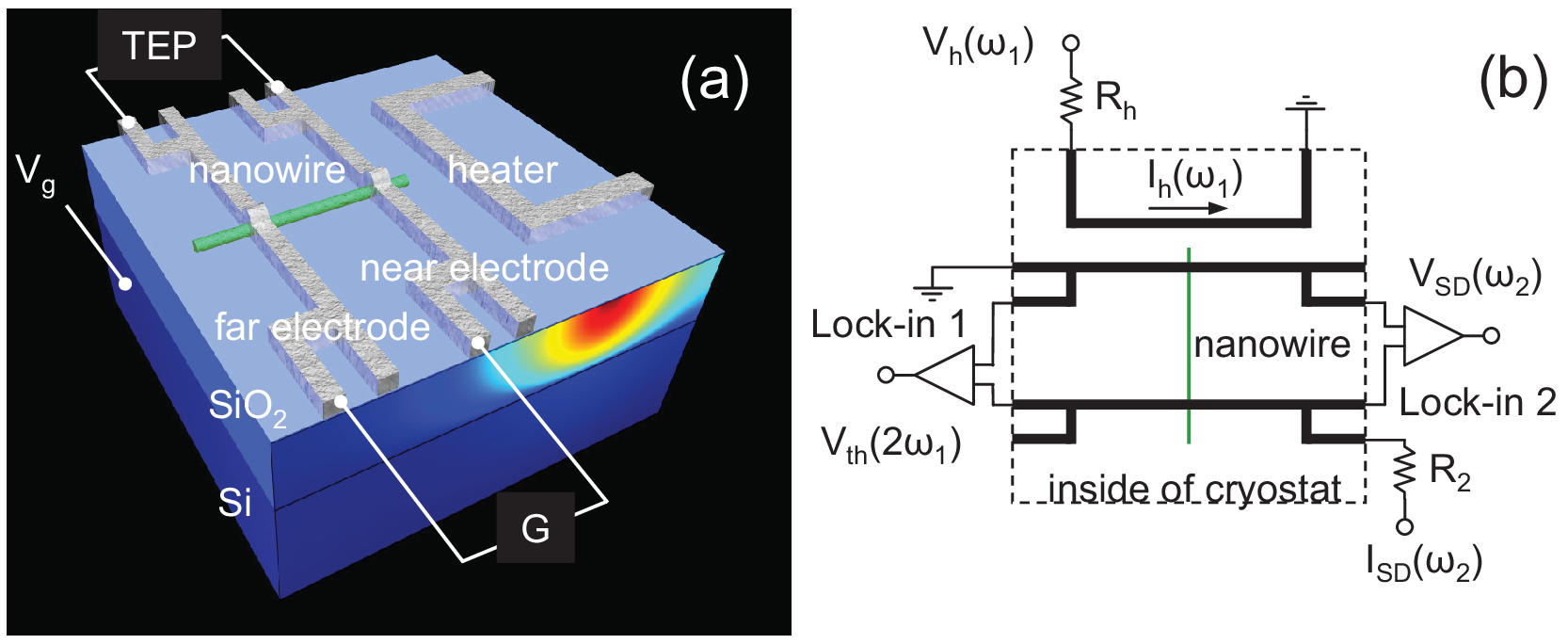}
\caption{(a) Schematic diagram of the simultaneous measurement technique of conductance and thermopower on individual nanowires. The finite element simulation shows a temperature profile, with red being the hottest and blue being the bath temperature, of the cross section of the substrate. (b) Circuit diagram of the AC configuration using 2 lock-in amplifiers. The TEP is measured at frequency $2\omega_1$ while the conductance is measured at frequency $\omega_2$.}
\end{figure}

In this letter we present an investigation of the EFE dependent the electronic and the thermoelectric transport properties of both Ge/Si core-shell NWs and Si NWs in the temperature range of 80-300 K. For this study, we first demonstrated the reliability of the EFE by simultaneous measuring both conductivity and TEP using two lock-in amplifiers, operating at two different excitation frequencies. In the individual nanowire FET device setting, we found a large modulation of TEP as a function of applied gate voltages induced by the EFE. Substantially large peak TEP values up to $\gtrsim 300~\mu$V/K are observed in the subthreshold regime of the Si and Ge/Si devices, indicating largely enhanced TEP near the band edge of semiconducting NWs.

The Si NWs, used in this study, were synthesized using the vapor-liquid-solid method (VLS) described in detail elsewhere~\cite{Si_schottky,Si_growth}. A typical diameters of the NWs are in the range of 20$\pm$5 nm and the axial orientation NWs are in the [110] direction. During the growth the NWs were doped with boron with a ratio of Si:B 8000:1. The Si NWs were subsequently suspended in ethanol and deposited onto a degenerately doped silicon substrate with 500 nm thermally grown SiO$_2$. The Si substrate back gate is capacitively coupled to the NW samples in order to modulate their carrier density with the EFE. Electron beam lithography, metallization (2/40 nm Ti/Pd), and liftoff procedure are used to define the heater and microthermometer structures. The samples are dipped into HF acid for 5 s immediately prior to metallization in order to remove native oxide. Another semiconducting system we employed in this study are core-shell heterostructured Ge/Si NWs. This heterostructured NWs were chosen since they are known to provide highly conductivity 1-dimensional hole gas at the core-shell interface~\cite{GeSi_PNAS}. The details of the synthesis of the core/shell Ge/Si heterostructure NWs, with diameters in the range of 12$\pm$2~nm, has been described previously~\cite{GeSi_PNAS}. The fabrication procedure of the FET-style devices for the TEP measurement was similar to that for the Si NWs, except that the electrodes were made from 50 nm Ni. The thick layer of Ni electrodes are employed to contact the 1-dimensional hole gas by rapid thermal annealing essential for the elimination of a Schottky barrier due to the diffusion of Ni through the Si shell layer.

\begin{figure}
\includegraphics[width=1.0\linewidth]{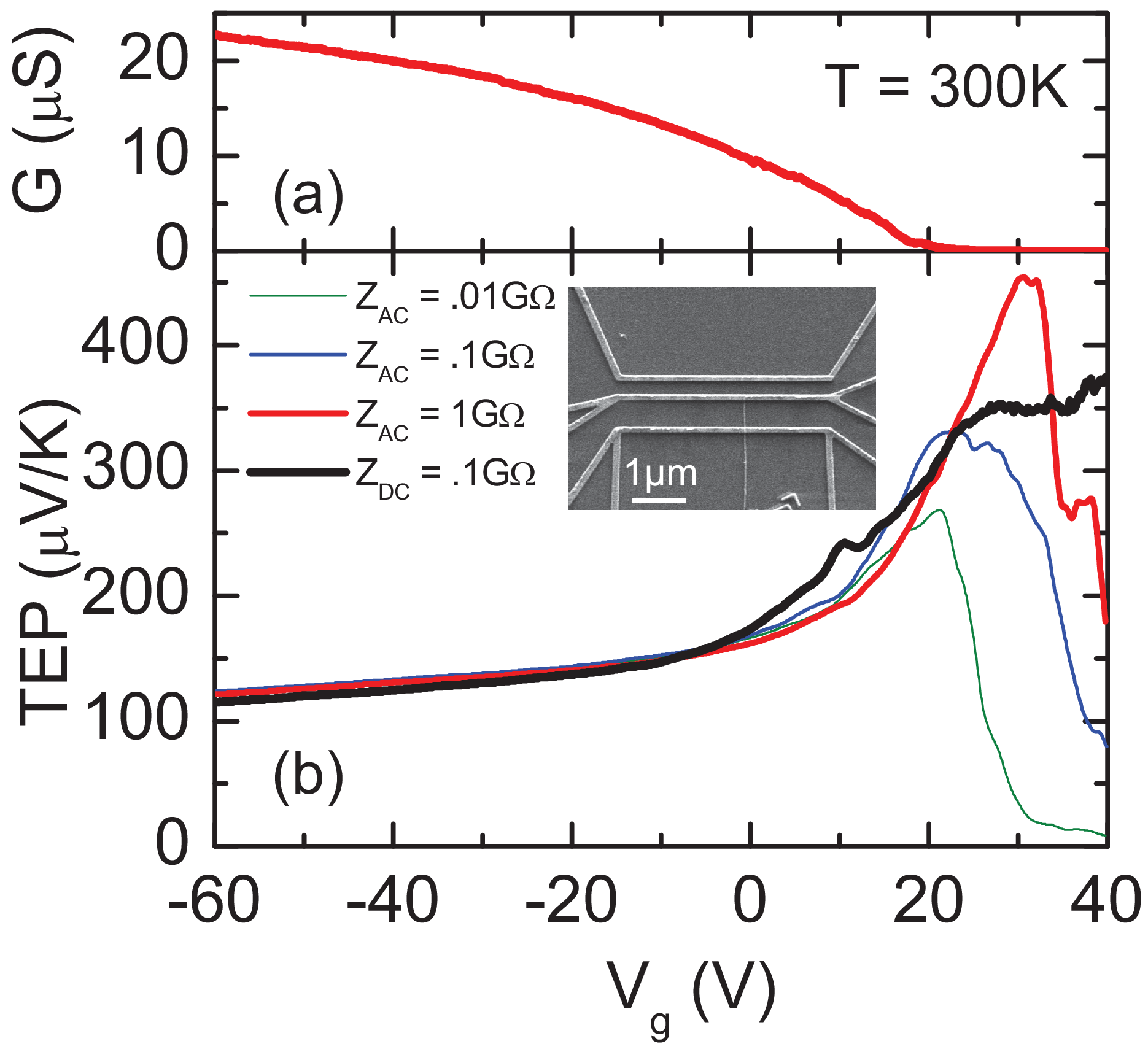}
\caption{ Conductance (a) and thermopower (b) of a Ge/Si nanowire as a function of gate voltage taken at $T=300$~K. The inset in (b) shows a typical SEM image of a 12~nm Ge/Si device. Large input impedance becomes important when measuring TEP near the band edge of a semiconductor, as the FET device turns off. }
\end{figure}

Conductance and TEP were measured in a vacuum cryostat, with pressure $\sim10^{-6}$ Torr. The technique to simultaneously measure conductance and TEP has been previously used to measure carbon nanotubes~\cite{CNT_TEP}, graphene~\cite{graphene_TEP}, and nanowires~\cite{SbTe_TEP, ZnO_TEP}. A schematic diagram and a circuit diagram are presented in Figure 1(a) and 1(b), respectively. Since SiO$_2$ is $\sim$100 times less thermally conductive than Si, the Joule heat generated by the heater electrode is mostly dissipated into the Si substrate, as seen in the finite element simulation (we used the software package COMSOL) in Figure 1(a). A lateral temperature gradient forms because the device geometry is chosen such that the separation between the heater and the near electrode is on the order of the SiO$_2$ thickness. Since resistance $R\propto T$, where $T$ is the temperature, in metals, the 4-probe electrodes act as microthermometers to measure the applied temperature difference, $\Delta T$. In the DC configuration, the thermally induced voltage $\Delta V_{th}$ was measured with a voltage amplifier to acquire the TEP, $S = -\frac{\Delta V_{th}}{\Delta T}$. The DC technique, however, is quite time consuming since at each gate voltage point the heater has to be swept in the wide range of the bias current to produce the appropriate temperature gradient.

In the AC configuration, shown in Figure 1(b), an AC voltage $V_h(\omega_1)$ is applied to the heater electrode. The temperature difference formed along the channel will be proportional to the square of that voltage, therefore the resulting voltage will oscillate at $2\omega_1$, with a $90^{\circ}$ phase shift. The TEP is then $S = -\frac{\sqrt{2} V_{th}(2\omega_1)}{\Delta T}$, where the $\sqrt{2}$ factor comes from the fact that the lock-in amplifier measures root-mean-squared values. For a consistency check we make sure that the DC and AC configurations produce the same TEP values. The condition of linear response, $\Delta T\ll T$, is always satisfied during the measurement in order to stay in the linear response regime. The conductance was measured using the standard 2-probe current biasing technique at $\omega_2$. Both signals are measured simultaneously as the carrier density is changed in the NW with applied gate voltage, $V_g$.

We first discuss the results from highly conductive Ge/Si NWs. A typical room temperature, $T = 300$~K, gate dependent conductance and TEP measurement of a Ge/Si NW is shown in Figure 2(a) and 2(b), respectively. A scanning electron microscope (SEM) image of a typical device is shown in the inset of Figure 2(b). Both conductance and TEP are modulated by applied gate voltage, $V_g$. For $V_g < 0$, the device exhibit high conduction, where two terminal conductance is in the order of ballistic conduction value $2e^2/h$, indicating high quality hole gas conduction. As a positive gate voltage is applied, the conductance decreases and the device turns off. This p-type behavior is expected from the 1D hole gas at the interface of the Ge/Si core/shell heterostructure. For negative gate voltages the TEP saturates to a constant value $\sim 120~\mu$V/K. As this FET device turns off the TEP begins to rise with a peak value of $\sim 350~\mu$V/K. Figure~2(b) shows the difference that instrument input impedance makes on measuring TEP when the Fermi level is near or inside the gap of a semiconductor. Near the turn-off regime, the impedance of the device becomes high and the voltage measurement across the channel becomes challenging. In order to investigate the effect of input impedance to the measured TEP values near the turn-off regime, we employed voltage preamplifiers with different DC impedance $Z_{DC}$ and AC impedance $Z_{AC}$ values, ranging from .01 G$\Omega$ to 1 G$\Omega$. We found that generally, the resistance of the NW channel $R$, defined from the 2-terminal conductance as $R =1/G$, is on the same order as the input impedance of the measurement instrument, the measured values of the TEP become unreliable. It was also found also that the measured TEP was independent of measurement frequency for $w_1< 100$~Hz.

The measurement of highly conducting Ge/Si NWs provide us general insight about the gate dependent TEP measurements in NW FET devices. In the degenerate regime, when the Fermi level is far away from a band edge, thermal equilibrium is established quickly on the timescale of the measurement frequency. However, near the band edge very few carriers participate in transport and the conductance is exponentially suppressed. The TEP rises when the Fermi level moves into the gap, however, the absence of adequate equilibration between the electrodes limits accurate measurement. In order to increase thermoelectric efficiency both $G$ and TEP have to be increased simultaneously, however, quite often the two parameters, as observed in our NWs, are inversely related~\cite{Macdonald}. Similar observation has been reported in recent work of less conductive Ge/Si core-shell NW devices~\cite{GeSi_TEP}.

\begin{figure}
\includegraphics[width=1.0\linewidth]{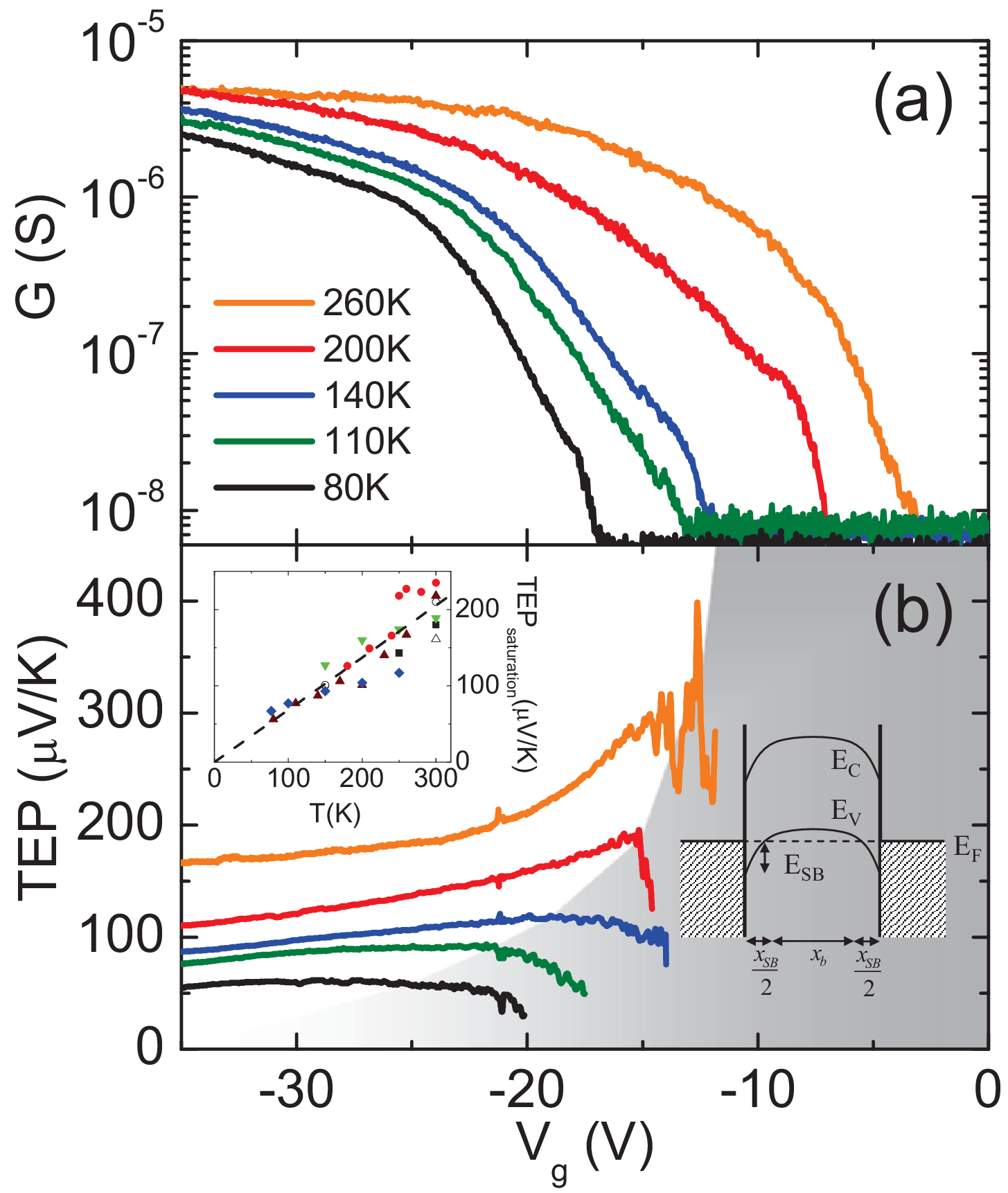}
\caption{ Conductance (a) and thermopower (b) of a 20 nm Si nanowire plotted as a function of gate voltage at (top to bottom) 260 K, 200 K, 140 K, 110 K, and 80 K. The shaded grey region in (b) is where the FET is off and the TEP values cannot be trusted. The upper inset of (b) shows a linear fit of the saturation TEP as a function of temperature from several devices. The lower inset of (b) shows an energy band diagram of the formation of a Schottky barrier in the Si NW - electrode system. }
\end{figure}

We now turn our attention to the Si NW measurements. Figure~3(a) shows our main result, the gate dependence of the conductance and TEP of a typical Si NW in the temperature range of 80-300~K. The FET is in the saturation regime for higher negative gate voltages. The height of the Schottky barrier $E_{SB}$ (see lower inset in Figure~3(b)) that forms between the Si NW and the metal electrode interface is adjusted by the applied gate voltage. In the subthreshold regime the conductance is drastically decreasing but still finite. In this regime, the mobility of the device is estimated from the transconductance $dG/dV_g$ to be $17$cm$^2$/Vs from $\mu = \frac{L^2}{C_g}\frac{dG}{dV_g}$ where $L$ is the device length and $C_g$ is the gate capacitance whose value can be estimated by the cylinder-on-plane model\cite{GeSi_PNAS}. Typical mobilities of measured samples vary between $0.1-20$~cm$^2$/Vs. The FET will be turned off at positive $V_g$ where the Schottky barriers deplete all available itinerant states in the valence band.

The total TEP is a contribution of the TEP from the bulk of the NW and from the Schottky barrier. At negative gate voltages, the Schottky barrier becomes very thin and contributes negligibly to the TEP, which saturates to a constant value at each temperature. This saturation TEP scales linearly as a function of temperature with a slope of 0.68$~\mu$V/K$^2$, as shown in the upper inset of Figure 3b, which signifies diffusive thermoelectric generation. In highly doped ($>10^{18}$~cm$^{-3}$) bulk silicon, TEP as a function of temperature is linear, while only lightly doped samples show non-monotonic behavior due to phonon-drag effects.\cite{Sibulk_TEP} The TEP is positive, which is consistent with the p-type nature of this material since the sign of the TEP indicates the carrier type.\cite{Ashcroft_Mermin} The carrier density can be extracted from the temperature dependence of the saturation TEP using the Mott relation:\cite{Cutler_Mott,ZnO_TEP}

\begin{equation}
S = \left. -\frac{\pi^2k_B^2T}{3|e|} \frac{d\ln \sigma}{dE}\right |_{E=E_F} = -\frac{\pi^2k_B^2m^*}{(3\pi^2)^{2/3}\hbar^2|e|}\frac{T}{n^{2/3}}
\end{equation}

where $k_B$ is the Boltzmann constant, $m^*$ is the effective mass, and $n$ is the carrier density. Using a parabolic dispersion relation with a hole effective mass\cite{Si_mass} of $m^*=0.39m_e$, the carrier density in the Si NWs is calculated to be $n \approx 1.4 \times 10^{-19}$cm$^{-3}$. Since the carrier concentration is a result of singly ionized B dopants, the ionized impurity concentration is $1.4 \times 10^{-19}$cm$^{-3}$. Because not all of the boron is converted during synthesis, the measured value of the impurity concentration is reasonable compared to the nominal value of $9\times 10^{20}$~cm$^{-3}$. As the Schottky barrier becomes wider in the subthreshold regime, thermally activated carriers hopping over the barrier contribute more to the overall measured TEP while carriers in the bulk contribute less. The peak values of the TEP at $T = 300$~K are $\sim 300~\mu$V/K. These values are higher than $220~\mu$V/K, measured in previous experiments of similar diameter Si NWs and doping levels in the suspended device geometry,\cite{SiNW_ZT1,SiNW_ZT2} however, are lower than values measured in the bulk.\cite{Sibulk_TEP}

In conclusion, we have measured the gate dependence of conductance and TEP of individual semiconducting p-type Si and Ge/Si NWs in the temperature range of 80-300~K. High input impedance is essential to measure TEP accurately when the FET is off. We have found peak TEP values of $300~\mu$V/K and $350~\mu$V/K in Si and Ge/Si NWs, respectively, in the subthreshold regime. The linear temperature dependence of the saturation TEP in Si NWs is used to acquire the dopant density to be $1.4 \times 10^{-19}$~cm$^{-3}$. Controlling the magnitude of the TEP using the EFE is essential in order to incorporate NWs into high efficiency thermoelectric power conversion devices.

This work was financially supported by the Korean Agency for Defense Development (ADD), under agreement number ADD-10-70-07-03.


\begin{thebibliography}{99}

\bibitem{Si_schottky}  
Y. Cui, X. Duan, J. Hu, and C. M. Lieber, J. Phys. Chem. B \textbf{104}, 5213 (2000).

\bibitem{Si_FET1}  
Y. Cui, Z. Zhong, D. Wang, W. U. Wang, and C. M. Lieber, Nano Lett. \textbf{3}, 149 (2003).

\bibitem{Si_FET2}  
G. Zheng, W. Lu, S. Jin, C. M. Lieber, Adv. Mater. \textbf{16}, 1890 (2004).

\bibitem{Si_growth}  
Y. Wu, Y. Cui, L. Huynh, C. J. Barrelet, D. C. Bell, and C. M. Lieber, Nano Lett. \textbf{4}, 433 (2004).

\bibitem{Si_mass}  
Z. Zhong, Y. Fang, W. Lu, and C. M. Lieber, Nano Lett. \textbf{5}, 1143 (2005).

\bibitem{Si_dopant}  
P. Xie, Y. Hu, Y. Fang, J. Huang, and C. M. Lieber, Proc. Natl. Acad. Sci. U.S.A. \textbf{106}, 15254 (2009).

\bibitem{GeSi_PNAS}  
W. Lu, J. Xiang, B. P. Timko, Y. Wu, and C. M. Lieber, Proc. Natl. Acad. Sci. U.S.A. \textbf{102}, 10046 (2005).

\bibitem{GeSi_FET}  
J. Xiang, W. Lu, Y. Hu, Y. Wu, H. Yan, and C. M. Lieber, Nature \textbf{441}, 489 (2006).

\bibitem{GeSi_theory}  
G. Liang, J. Xiang, N. Kharche, G. Klimeck, C. M. Lieber, and M. Lundstrom, Nano Lett. \textbf{7}, 642 (2007).

\bibitem{GeSi_2THz}  
Y. Hu, J. Xiang, G. Liang, H. Yan, and C. M. Lieber, Nano Lett. \textbf{8}, 925 (2008).

\bibitem{SiNW_ZT1}  
A. I. Hochbaum, R. Chen, R. D. Delgado, W. Liang, E. C. Garnett, M. Najarian, A. Majumdar, and P. Yang, Nature \textbf{451}, 163 (2007).

\bibitem{SiNW_ZT2}  
A. I. Boukai, Y. Bunimovich, J. Tahir-Kheli, J. K. Yu, W. A. Goddard III, and J. R. Heath, Nature \textbf{451}, 168 (2007).

\bibitem{PbSe_FET}  
W. Liang, A. I. Hochbaum, M. Fardy, O. Rabin, M. Zhang, and P. Yang, Nano Lett. \textbf{9}, 1689 (2009).

\bibitem{SbTe_TEP}  
Y. Zuev, J. S. Lee, C. Galloy, H. Park, and P. Kim, Nano Lett. \textbf{12}, 3037 (2012).

\bibitem{InAs_TEP}  
Y. Tian, M. R. Sakr, J. M. Kinder, D. Liang, M. J. MacDonald, R. L. J. Qiu, H.-J. Gao, and X. P. A. Gao, Nano Lett. \textbf{12}, 6492 (2012).

\bibitem{GeSi_TEP}  
J. Moon, J.-H. Kim, Z. C. Y. Chen, J. Xiang, and R. Chen, Nano Lett. \textbf{13}, 1196 (2013).

\bibitem{CNT_TEP}  
J. Small, K. Perez, and P. Kim, Phys. Rev. Lett. \textbf{91} 256801, (2003).

\bibitem{graphene_TEP}  
Y. M. Zuev, W. Chang, and P. Kim, Phys. Rev. Lett. \textbf{102} 096807, (2009).

\bibitem{ZnO_TEP}  
C. Lee, G. Yi, Y. M. Zuev, and P. Kim, Appl. Phys. Lett. \textbf{94} 022106, (2009).

\bibitem{Ashcroft_Mermin}  
N. W. Ashcroft and N. D. Mermin, Solid State Physics (Thomson Learning, Inc., USA, 1976).

\bibitem{Macdonald}  
D. K. C. Macdonald, Thermoelectricity (Dover, New York, 2006).

\bibitem{Cutler_Mott}  
M. Cutler and N. F. Mott, Phys. Rev. \textbf{181} 1336, (1969).

\bibitem{Sibulk_TEP}  
T. H. Geballe and G. W. Hull, Phys. Rev. \textbf{98} 940, (1955).



\end{thebibliography}
\end{document}